\documentstyle[12pt]{article}
\newcommand{\ssigma}{\hbox{$\kern2.5pt\vrule height4pt\kern-2.5pt\sigma$}}
\newcommand{\oone}{\hbox{$1\kern-2pt{\rm l}\kern2pt$}}

\newcommand{\GeV}{{\sl\,GeV}}
\newcommand{\real}{{\sl Re\,}}

\newcommand{\GG}{{\cal G}}
\newcommand{\GL}{{\cal G}_\ell}
\newcommand{\Li}{{\rm Li}_2}
\begin{document}

\thispagestyle{empty}
\begin{flushright}
MZ-TH/97-17\\
hep-ph/9806464\\
June 1998
\end{flushright}
\vspace{0.5cm}
\begin{center}
{\Large\bf Gluon Polarization in 
  {\boldmath$e^+e^-\rightarrow t\bar tG$}:}\\[.3truecm]
{\Large\bf Polar Angle Dependence and}\\[.3truecm]
{\Large\bf Beam Polarization Effects}\\[1.3truecm]
{\large S.~Groote, J.G.~K\"orner and J.A.~Leyva\footnote{on leave absence 
from CIF, Colombia}}\\[1truecm]
Institut f\"ur Physik, Johannes-Gutenberg-Universit\"at,\\[.2cm]
Staudinger Weg 7, D-55099 Mainz, Germany\\
\end{center}
\vspace{1cm}
\begin{abstract}\noindent
We calculate the linear and circular polarization of gluons produced in 
conjunction with massive quarks in the annihilation process
$e^+e^-\rightarrow q \bar q G$. The linear polarization is calculated in the 
hadron event plane as well as in the gluon-beam plane. Beam polarization
and polar orientation effects are included in our discussion. For typical 
top pair production energies at the Next-Linear-Collider (NLC) the degree 
of linear polarization in the hadron event plane remains close to its soft 
gluon value of 100\% over most of the energy spectrum of the gluon. The 
linear polarization in the gluon-beam plane is generally smaller but peaks 
toward the hard end of the gluon spectrum. The dependence of the linear 
polarization on beam polarization and on the polar orientation of the 
gluon is small. The circular polarization is largest for maximal gluon 
energies and shows a strong dependence on the longitudinal beam 
polarization. The longitudinal polarization of the beam may therefore be 
used to tune the circular polarization of the gluon. The massive quark 
results are compared with the corresponding results for the massless quark 
case.
\end{abstract}

\vspace{0.5cm}

\newpage

\section{Introduction}
The polarization of gluons in $e^+e^-$ annihilation~\cite{gluon1,gluon2}, 
in deep inelastic scattering~\cite{gluon3} and in quarkonium 
decays~\cite{gluon1,gluon4} has been studied in a series of papers dating 
back to the early 1980's. Several proposals have been put forward to measure
the polarization of the gluon among which is the proposal to measure
azimuthal angular correlation effects in the splitting process of a 
polarized gluon into a pair of gluons or quarks~\cite{gluon5}. Latter 
proposal has led to a beautiful confirmation of the presence of the 
three-gluon vertex using $e^+e^-$ data~\cite{gluon6} (see also~\cite{gluon7}).

The earlier calculation of the gluon's polarization in $e^+e^-$ 
annihilations had been done for massless fermions which was quite 
sufficient for the purposes of that period. In the meantime the situation 
has changed in as much as the heavy top quark has been discovered at 
Fermilab in 1995 whose production properties in $e^+e^-$ annihilations will 
be studied in the proposed Next-Linear-Collider (NLC). In its first stage 
typical running energies of the NLC would extend from $t\bar t$-threshold 
at about $350\GeV$ to maximal energies of about $550\GeV$. It is quite 
clear that top mass effects cannot be neglected in this energy range even 
at the highest c.m.~energies. It is therefore timely to redo the 
calculations of~\cite{gluon1,gluon2} for heavy quarks and to investigate 
the influence of heavy quark mass effects on the polarization observables 
of the gluon. A first step in this direction was taken by us 
in~\cite{gluon8} where we determined the linear polarization of the gluon 
in the process $e^+e^-\rightarrow t\bar tG$ taking the hadron event plane
(for short: event plane) as a reference plane. In~\cite{gluon8} we did not 
take into account azimuthal and polar orientation effects of the event 
plane relative to the beam. This is done by an appropriate polar and 
azimuthal averaging process. Beam polarization effects were only commented 
on in passing in~\cite{gluon8}.

In this paper we extend the analysis of~\cite{gluon8} in several directions.
We include beam-event orientation and beam polarization effects in our
discussion. We also compute the circular polarization of the gluon induced 
by the parity-odd component of the hadron tensor and/or by longitudinal 
beam polarization effects. Finally we compute the linear polarization of 
the gluon in the gluon-beam plane which is obtained after an appropriate 
azimuthal averaging process.

Our paper is structured as follows. Sec.~2 describes the general formalism
of massive quark plus gluon production in $e^+e^-$ annihilations including
beam and gluon polarization effects. Our production cross section is
written in terms of three modular building blocks. The first building
block defines the orientation of the lepton beam relative to the hadron
plane and the dependence on the polarization parameters of the beam.
The second building block specifies the electro-weak model dependence and
the third building block specifies the QCD dynamics in terms of a set of
polarized and unpolarized structure functions. In Sec.~3 we list our results
for the twice-differential $O(\alpha_s)$ polarized and unpolarized structure
functions as well as closed form expressions for their once- and
twice-integrated forms. Sec.~4 contains our numerical results. We provide 
plots of the energy dependence of the linear and circular polarization of 
the gluon, their polar angle dependence and their beam polarization 
dependence. In Sec.~5 we discuss the linear polarization of the gluon in 
the gluon-beam plane which can be obtained via an azimuthal rotation from 
the event plane. Sec.~6 contains our summary and our conclusions.
 
Before we get to the main topic of this paper we want to briefly present
some numerical evidence for what is referred to as the ``dead-cone'' effect
which occurs when soft gluons are radiated off heavy quarks. In the heavy 
quark case there is a depletion of gluons close to the direction of the 
heavy quark and the antiquark. This is quite different from the case of 
gluons radiated off light quarks where the production peaks towards the 
collinear limit. Large angle emission is a welcome effect in $e^+e^-$ 
annihilation since gluons emitted at large angles from the heavy quark 
and antiquark directions are easier to reconstruct. If one is far enough
away from the production threshold with sufficient velocities of the heavy 
quark and antiquark the decay products of the heavy quark and antiquark 
would tend to go along the original production direction and would thus 
stay away from the large-angle soft gluons radiated from the heavy quark 
and antiquark. For example, at $\sqrt{q^2}=500\GeV$ and 
$\sqrt{q^2}=1000\GeV$ one has a quark (or antiquark) velocity of $v=0.71$ 
and $v=0.94$, respectively.

In order to display the ``dead-cone'' effect we have plotted the quark/gluon
opening angle distribution at a c.m.\ energy of $\sqrt{q^2}=500\GeV$ for 
four different soft gluon energies in Fig.~1(a). The quark gluon opening 
angle $\theta_{13}$ is given by
\begin{equation}
\cos \theta_{13}= \frac {x^2+(2-x)w}{x\sqrt{(2-x+w)^2-4\xi}}
\end{equation}
At these low energies 
there is in fact a depletion of gluons radiated at small opening angles. 
However, the distributions rise uniformly to their maximum values at $90^0$ 
and do not show a peak at around $\theta_{13}=2m/\sqrt{q^2}=40.10^0$ as 
predicted in~\cite{gluon10}. In Fig.~1(b) we show the opening angle 
distribution for gluons at a fixed energy of $E=6\GeV$ for three different 
values of the c.m.\ energy. Again there is a depletion of gluons radiated 
at small opening angles for all three c.m.\ energies. The 
$\sqrt{q^2}=1000\GeV$ opening angle distribution does show a peak at around
$\theta_{13}=40^0$ which is somewhat displaced from the peak position
$\theta_{13}=20.05^0$ as calculated from the above relation. The two 
distributions at lower c.m.\ energies do not show any peak structure but 
rise uniformly to their maximum values at $90^0$.

We mention that some indirect evidence for the ``dead-cone'' effect was 
presented in~\cite{gluon11} where we calculated the mean transverse 
momentum of the top with regard to the antitop direction in $e^+e^-$ 
annihilations. The mean transverse momentum of the top was found to be 
only slightly below the mean gluon energy in the whole range from
$t\bar t$-threshold to $1000\GeV$. Since the transverse momentum of the 
top has to be balanced by the transverse momentum of the gluon the near 
equality of the two means implies large mean opening angles for the gluon.

\section{General formalism} 
As is usual we shall represent the two-by-two differential density matrix 
$d\ssigma=d\sigma_{\lambda_G\lambda'_G}$ of the gluon with gluon helicities 
$\lambda_G=\pm 1$ in terms of its components along the unit matrix and the 
three Pauli matrices. Accordingly one has
\begin{equation}\label{eqn1}
d\ssigma=\frac12(d\sigma\oone+d\sigma^x\ssigma_x+d\sigma^y\ssigma_y
  +d\sigma^z\ssigma_z),
\end{equation}
where $d\sigma$ is the unpolarized differential rate and 
$d\vec\sigma=(d\sigma^x,d\sigma^y,d\sigma^z)$ are the three components of 
the (unnormalized) differential Stokes vector. In this paper the
$(x,y,z)$-components of the Stokes vector will mostly be referred to the
event plane such that the gluon points in the $z$-direction and the
$(q\bar q G)$-plane defines the $(x,z)$-plane (for short: event plane). An 
exception is Sec.~5 where the linear polarization of the gluon is 
calculated in the gluon-beam plane.

Specifying to $e^+e^-\rightarrow q(p_1)\bar q(p_2)G(p_3)$ we perform an 
azimuthal averaging over the relative beam-event orientation. After 
azimuthal averaging the $y$-component of the Stokes vector $d\sigma^y$ 
drops out~\cite{gluon1,gluon2}\footnote{The linear gluon-beam plane
polarization observable of the gluon discussed in Sec.~5 involves a 
different azimuthal averaging process. For this application one needs to 
retain one particular $y$-component of the Stokes vector in the event 
plane.}. One retains only the $x$- and $z$-components of the Stokes vector 
which are referred to as the gluon's linear polarization in the event 
plane and the circular polarization of the gluon, respectively. The 
differential unpolarized and polarized rates, differential with regard to 
the polar beam-event orientation and the two energy-type variables 
$x=2p_3\cdot q/q^2$ and $w=2(p_1-p_2)\cdot q/q^2$ (with $q=p_1+p_2+p_3$) 
are then given by
\begin{eqnarray}
\frac{d\sigma^{(x)}}{d\cos\theta\,dx\,dw}
  &=&\frac38(1+\cos^2\theta)\left(g_{11}\frac{d\sigma_U^{1(x)}}{dx\,dw}
  +g_{12}\frac{d\sigma_U^{2(x)}}{dx\,dw}\right)\label{eqn2}\\&&
  +\frac34\sin^2\theta\left(g_{11}\frac{d\sigma_L^{1(x)}}{dx\,dw}
  +g_{12}\frac{d\sigma_L^{2(x)}}{dx\,dw}\right)
  +\frac34\cos\theta\,g_{44}\frac{d\sigma_F^{4(x)}}{dx\,dw},\nonumber\\
\frac{d\sigma^z}{d\cos\theta\,dx\,dw}
  &=&\frac38(1+\cos^2\theta)g_{14}\frac{d\sigma_U^{4z}}{dx\,dw}
  +\frac34\cos\theta\left(g_{41}\frac{d\sigma_F^{1z}}{dx\,dw}
  +g_{42}\frac{d\sigma_F^{2z}}{dx\,dw}\right).\qquad\label{eqn3}
\end{eqnarray}
The notation $d\sigma^{(x)}$ stands for either $d\sigma$ or $d\sigma^x$, 
and the same for $d\sigma_\alpha^{i(x)}$ (the indices $i=1,2,4$ and
$\alpha=U,L,F$ are explained later on). The notation closely follows the 
one used in~\cite{gluon6}. Noteworthy is the absence of a longitudinal 
contribution to the circular polarization with an angular 
$\sin^2\theta$-dependence. This is a tree-level effect due to the 
$CP$-evenness of the Standard Model interactions. 

We have written the electro-weak cross section in modular form in terms of
three building blocks. The first building block determines the angular
beam-event dependence. For the case at hand one remains with a polar angle
dependence after azimuthal integration, where $\theta$ is the polar angle
between the gluon and the electron beam. The second building block specifies
the electro-weak model dependence through the parameters $g_{ij}$ 
($i,j=1,\ldots,4$). They are given by
\begin{eqnarray}
g_{11}&=&Q_f^2-2Q_fv_ev_f\real\chi_Z+(v_e^2+a_e^2)(v_f^2+a_f^2)|\chi_Z|^2,
  \nonumber\\
g_{12}&=&Q_f^2-2Q_fv_ev_f\real\chi_Z+(v_e^2+a_e^2)(v_f^2-a_f^2)|\chi_Z|^2,
  \nonumber\\
g_{14}&=&2Q_fv_ea_f\real\chi_Z-2(v_e^2+a_e^2)v_fa_f|\chi_Z|^2,\\
g_{41}&=&2Q_fa_ev_f\real\chi_Z-2v_ea_e(v_f^2+a_f^2)|\chi_Z|^2,\nonumber\\
g_{42}&=&2Q_fa_ev_f\real\chi_Z-2v_ea_e(v_f^2-a_f^2)|\chi_Z|^2,\nonumber\\
g_{44}&=&-2Q_fa_ea_f\real\chi_Z+4v_ea_ev_fa_f|\chi_Z|^2,\nonumber
\end{eqnarray}
where, in the Standard Model, 
$\chi_Z(q^2)=gM_Z^2q^2/(q^2-M_Z^2+iM_Z\Gamma_Z)$, with $M_Z$ and $\Gamma_Z$ 
the mass and width of the $Z^0$ and 
$g=G_F(8\sqrt2\pi\alpha)^{-1}\approx 4.49\cdot 10^{-5}\GeV^{-2}$. $Q_f$ are 
the charges of the final state quarks; $v_e$ and $a_e$, $v_f$ and $a_f$ are 
the electro-weak vector and axial vector coupling constants. For example, 
in the Weinberg-Salam model, one has $v_e=-1+4\sin^2\theta_W$, $a_e=-1$ for 
leptons, $v_f=1-\frac83\sin^2\theta_W$, $a_f=1$ for up-type quarks 
($Q_f=\frac23$), and $v_f=-1+\frac43\sin^2\theta_W$, $a_f=-1$ for down-type 
quarks ($Q_f=-\frac13$). In this paper we use Standard Model couplings with 
$\sin^2\theta_W=0.226$.

The third building block, finally, is given by the hadron dynamics, i.e.\ 
by the current-induced production of a heavy quark pair with subsequent 
gluon emission. The QCD dynamics is encoded in the hadronic rate 
functions $d\sigma_\alpha^{i(x)}/dx\,dw$ and $d\sigma_\alpha^{iz}/dx\,dw$ 
with specific components denoted by the indices $\alpha$ and $i$, and the
polarization indices $x$ and $z$. The hadronic rate functions can in turn 
be related to the polarized and unpolarized hadronic structure functions 
$H_\alpha^{i(x)}$ and $H_\alpha^{iz}$ according to
\begin{equation}
\frac{d\sigma_\alpha^{i(x)}}{dx\,dw}
  =\frac{\alpha^2}{6\pi q^2}H_\alpha^{i(x)}(x,w),\qquad
\frac{d\sigma_\alpha^{iz}}{dx\,dw}
  =\frac{\alpha^2}{6\pi q^2}H_\alpha^{iz}(x,w).
\end{equation}
The different hadronic structure functions are specific components of the
hadronic three-body tensor $H_{\mu \nu}$ induced by the product of vector 
current ($V$) and axial vector current ($A$) interactions. In the case
of massive quark pair production one has four different independent current 
products which are denoted by the upper index $i=1,\ldots,4$, where 
\begin{eqnarray}\label{eqna}
H^1=\frac12(H^{VV}+H^{AA}),&&H^2=\frac12(H^{VV}-H^{AA}),\nonumber\\
H^3=\frac i2(H^{V\!A}-H^{AV}),&&H^4=\frac12(H^{V\!A}+H^{AV}).
\end{eqnarray}
We have temporarilly dropped all further indices on the hadron tensor
in Eq.~(\ref{eqna}).

We mention that, because of $CP$-invariance and because we are calculating 
at tree level, the linear combination $H^3$ does not contribute to the 
cross section and the spin observables considered in this paper. The lower 
index $\alpha=U,L,F$ specifies the relevant density matrix element of the 
gauge boson that determines the polar angle $\cos\theta$-distribution. The 
three elements of the density matrix are referred to as the ``unpolarized 
transverse'' component $H_U=H_{++}+H_{--}=H_{11}+H_{22}$, the 
``longitudinal'' component $H_L=H_{00}$ and the ``forward-backward 
asymmetric'' component $H_F=H_{++}-H_{--}=-i(H_{12}-H_{21})$ where we have 
listed both the spherical and Cartesian components of the relevant density 
matrix elements. The three ($U,L,F$) components can be obtained from the 
full hadron tensor $H_{\mu\nu}$ by covariant contraction as discussed 
in~\cite{gluon6}. The unpolarized hadronic tensor $H_\alpha^i$ and the 
polarized hadronic tensors $H_\alpha^{ix}$ and $H_\alpha^{iz}$ are defined 
in exact analogy to Eq.~(\ref{eqn1}).

Eqs.~(\ref{eqn2}) and~(\ref{eqn3}) give the differential cross section for 
unpolarized beams. The case of longitudinally polarized beams can easily be 
included. For the unpolarized and linearly polarized rates
$d\sigma_\alpha^{i(x)}$ one has to effect the replacement
\begin{eqnarray}
g_{1i}&\rightarrow&(1-h^-h^+)g_{1i}+(h^--h^+)g_{4i}\qquad(i=1,2)\nonumber\\
g_{44}&\rightarrow&(h^--h^+)g_{14}+(1-h^-h^+)g_{44}
\end{eqnarray}
where $h^+$ and $h^-$ are the degrees of the longitudinal (or helicity) 
polarization of the positron and electron beam, respectively. For the 
circularly polarized rates $d\sigma_\alpha^{iz}$ one has the replacements
\begin{eqnarray}
g_{14}&\rightarrow&(1-h^-h^+)g_{14}+(h^--h^+)g_{44}\nonumber\\
g_{4i}&\rightarrow&(h^--h^+)g_{1i}+(1-h^-h^+)g_{4i}\qquad(i=1,2).
\end{eqnarray}
Transverse beam polarization effects can also be easily included in the
framework of our formalism (see e.g.~\cite{gluon9}) but will not be 
further discussed in this paper.

After having described the general formalism we now go on to present
explicit expressions for the polarized and unpolarized hadronic structure
functions for the annihilation process $e^+ e^-\rightarrow q \bar q G$
in the following section.

\section {Polarized and unpolarized structure functions}
The various components of the hadronic tensor can be easily calculated 
from the relevant tree level Feynman diagrams and are given by
\begin{eqnarray}
H_U^1(x,w)&=&-16\Big(\frac2x-2+x\Big)\frac1x
  -64t_+(x,w)+32\Big(\frac2x-2+x\Big)t_+^\ell(x,w),\nonumber\\
H_U^2(x,w)&=&-32\xi t_+(x,w)+16\xi\Big(\frac{2-\xi}x-2\Big)t_+^\ell(x,w),
  \nonumber\\
H_L^1(x,w)&=&32\Big(\frac1x-1\Big)\frac1x
  +16\xi t_+(x,w)-8\xi\Big(\frac{6-\xi}x-2-x\Big)t_+^\ell(x,w),\label{eqn4}\\
H_L^2(x,w)&=&-16\xi t_+(x,w)+8\xi\Big(\frac{2-\xi}x-2-x\Big)t_+^\ell(x,w),
  \nonumber\\
H_F^4(x,w)&=&64t_-(x,w)-32\Big(\frac2x-2+x\Big)t_-^\ell(x,w),\nonumber\\[12pt]
H_U^{1x}(x,w)&=-&32\Big(\frac1x-1\Big)\frac1x
  -64t_+(x,w)+64\Big(\frac1x-1\Big)t_+^\ell(x,w),\nonumber\\
H_U^{2x}(x,w)&=&-32\xi t_+(x,w)+16\xi\Big(\frac{2-\xi}x-2\Big)t_+^\ell(x,w),
  \nonumber\\
H_L^{1x}(x,w)&=&32\Big(\frac1x-1\Big)\frac1x
  +16\xi t_+(x,w)-8\xi\Big(\frac{6-\xi}x-2\Big)t_+^\ell(x,w),\\
H_L^{2x}(x,w)&=&-16\xi t_+(x,w)+8\xi\Big(\frac{2-\xi}x-2\Big)t_+^\ell(x,w),
  \nonumber\\
H_F^{4x}(x,w)&=&64t_-(x,w)-64\Big(\frac1x-1\Big)t_-^\ell(x,w),
  \nonumber\\[12pt]
H_U^{4z}(x,w)&=&64xt_-(x,w)-32(2-x)t_-^\ell(x,w),\nonumber\\[7pt]
H_L^{4z}(x,w)&=&0,\nonumber\\
H_F^{1z}(x,w)&=&-16(2-x)\frac1x-64xt_+(x,w)+32(2-x)t_+^\ell(x,w),\label{eqn5}\\
H_F^{2z}(x,w)&=&0\nonumber
\end{eqnarray}
with $\xi=4m_q^2/q^2$, where we have used the abbreviations
\begin{equation}\label{eqn6}
t_\pm(x,w):=\frac\xi4\left(\frac1{(x-w)^2}\pm\frac1{(x+w)^2}\right),\qquad
t_\pm^\ell(x,w):=\frac12\left(\frac1{x-w}\pm\frac1{x+w}\right).
\end{equation}
We mention that the vanishing of the hadron tensor components 
$H_L^{4z}(x,w)$ and $H_F^{2z}(x,w)$ is a tree-level effect and is due to 
the $CP$-invariance of the underlying Standard Model interaction.

Note that under quark-antiquark exchange (charge conjugation) 
$t_\pm\rightarrow\pm t_\pm$ and $t_\pm^\ell\rightarrow\pm t_\pm^\ell$. This 
implies that the charge conjugation odd contributions $t_-$ and $t_-^\ell$ 
vanish if one does not discriminate between quarks 
and antiquarks. In the following we wish to distinguish between the cases 
where the quark flavours are identified or not identified and we will refer 
to these two cases as the flavour tag and flavour no-tag cases, 
respectively.

Let us briefly pause to discuss the mass zero limit of the hadronic tensor 
components. The mass zero limit can easily be taken by setting $\xi=0$ in 
Eqs.~(\ref{eqn4}--\ref{eqn5}). From the terms that remain after taking the 
$\xi\rightarrow 0$ limit it is only the terms proportional to 
$t_\pm^\ell(x,w)$ that are important, because they are mass singular in the 
$\xi\rightarrow 0$ limit. In fact, when one performs the $w$-integration 
including flavour-tagging, the mass singluar functions $t_\pm^\ell(x,w)$ 
integrate to an $x$-independent logarithmic factor and a finite 
($x$-dependent) term, i.e.\ 
$\int t_\pm^\ell(x,w)dw\rightarrow-\ln\xi+c_\pm$ (see Eq.~(\ref{eqn6})). 
Keeping only the dominant logarithmic term one has in the $\xi\rightarrow 0$ 
limit
\begin{eqnarray}\label{eqn7}
H_U^1(x)\rightarrow-32\Big(\frac2x-2+x\Big)\ln\xi,&&
H_F^4(x)\rightarrow32\Big(\frac2x-2+x\Big)\ln\xi,\nonumber\\
H_U^{1x}(x)\rightarrow-64\Big(\frac1x-1\Big)\ln\xi,&&
H_F^{4x}(x)\rightarrow64\Big(\frac1x-1\Big)\ln\xi,\\
H_U^{4z}(x)\rightarrow32(2-x)\ln\xi,&&
H_F^{1z}(x)\rightarrow-32(2-x)\ln\xi.\nonumber
\end{eqnarray}
The remaining terms in Eqs.~(\ref{eqn4}--\ref{eqn5}) are subdominant. The 
expressions in Eqs.~(\ref{eqn7}) determine the energy dependence of the 
linear and circular polarization of the gluon in the mass zero limit. They 
will be used later on to compare with the corresponding expressions in the 
massive case.

Returning to the case of massive quarks, the integration over the 
phase-space parameter $w$ can be easily done. However, as mentioned
earlier on, $C$-odd observables average out to zero in this integration if 
one does not employ flavour-tagging. For example, the contributions to the 
circular polarization vanish when we integrate over the whole range of $w$ 
since the integration variable is antisymmetric in the quark and antiquark 
variables. Nonvanishing values of the circular polarization are obtained 
only if one applies flavour-tagging, i.e.\ one has to take care to 
distinguish between quark and antiquark energies in the integration. Thus,
for a fixed value of the gluon's energy, one first integrates over
positive values of the $w$-variable and then subtracts the integral over
negative $w$-values. Or, equivalently, one just takes twice the
value of the circular polarization calculated for positive $w$. 

Let us list the few basic $w$-integrals that are needed in the calculation.
The integration has to be done between the symmetric phase space boundaries 
$-w_+$ and $w_+$ where
\begin{equation}
w_+(x)=x\sqrt{\frac{1-x-\xi}{1-x}}.
\end{equation}
One has
\begin{eqnarray}
\int_{-w_+(x)}^{+w_+(x)}t_+(x,w)dw&=&\Big(\frac1x-1\Big)\frac1xw_+(x),
  \nonumber\\
\int_{-w_+(x)}^{+w_+(x)}t_+^\ell(x,w)dw&=&\ln\left(\frac{\sqrt{1-x}
  +\sqrt{1-x-\xi}}{\sqrt{1-x}-\sqrt{1-x-\xi}}\right)\ =:\ t_+^\ell(x),\\
2\int_0^{+w_+(x)}t_-(x,w)dw&=&\frac{1-\xi}x-1,\nonumber\\
2\int_0^{+w_+(x)}t_-^\ell(x,w)dw&=&\ln\left(\frac{1-x}\xi\right)
  \ =:\ t_-^\ell(x).\nonumber
\end{eqnarray}
where we have employed flavour-tagging for the last two $C$-odd integrals as
discussed before. Using these basic integrals we obtain the once-integrated
structure functions $H_\alpha^i(x)$, $H_\alpha^{ix}(x)$ and
$H_\alpha^{iz}(x)$. One has

\begin{eqnarray}
H_U^1(x)&=&-32\Big(\frac4x-4+x\Big)\frac1xw_+(x)
  +32\Big(\frac2x-2+x)t_+^\ell(x),\nonumber\\
H_U^2(x)&=&-32\xi\Big(\frac1x-1\Big)\frac1xw_+(x)
  +16\xi\Big(\frac{2-\xi}x-2\Big)t_+^\ell(x),\nonumber\\
H_L^1(x)&=&16(4+\xi)\Big(\frac1x-1\Big)\frac1xw_+(x)
  -8\xi\Big(\frac{6-\xi}x-2-x\Big)t_+^\ell(x),\\
H_L^2(x)&=&-16\xi\Big(\frac1x-1\Big)\frac1xw_+(x)
  +8\xi\Big(\frac{2-\xi}x-2-x\Big)t_+^\ell(x),\nonumber\\
H_F^4(x)&=&64\Big(\frac{1-\xi}x-1\Big)
  -32\Big(\frac2x-2+x\Big)t_-^\ell(x),\nonumber\\[12pt]
H_U^{1x}(x)&=&-128\Big(\frac1x-1\Big)\frac1xw_+(x)
  +64\Big(\frac1x-1)t_+^\ell(x)\nonumber\\
H_U^{2x}(x)&=&-32\xi\Big(\frac1x-1\Big)\frac1xw_+(x)
  +16\xi\Big(\frac{2-\xi}x-2\Big)t_+^\ell(x),\nonumber\\
H_L^{1x}(x)&=&16(4+\xi)\Big(\frac1x-1\Big)\frac1xw_+(x)
  -8\xi\Big(\frac{6-\xi}x-2\Big)t_+^\ell(x),\\
H_L^{2x}(x)&=&-16\xi\Big(\frac1x-1\Big)\frac1xw_+(x)
  +8\xi\Big(\frac{2-\xi}x-2\Big)t_+^\ell(x),\nonumber\\
H_F^{4x}(x)&=&64\Big(\frac{1-\xi}x-1\Big)
  -64\Big(\frac1x-1\Big)t_-^\ell(x),\nonumber\\[12pt]
H_U^{4z}(x)&=&64(1-\xi-x)-32(2-x)t_-^\ell(x),\nonumber\\[7pt]
H_L^{4z}(x)&=&0,\nonumber\\[7pt]
H_F^{1z}(x)&=&-32(4-3x)\frac1xw_+(x)+32(2-x)t_+^\ell(x),\\[7pt]
H_F^{2z}(x)&=&0.\nonumber
\end{eqnarray}
Our notation is such that $H(x)=\int H(x,w) dw$.
The dominant leading logarithmic $\ln\xi$ terms in the zero mass limit
derive from the logarithmic functions $t^l_+(x)$ and $t^l_-(x)$ and have
been listed before in Eq.~(\ref{eqn7}).

The last step is the integration over the second phase-space parameter $x$. 
It is clear that we have to introduce a gluon energy cutoff at the soft end 
of the gluon spectrum in order to keep the rate finite. Denoting the cutoff 
energy by $E_c=\lambda \sqrt{q^2}$ the integration extends from 
$x=2\lambda=2E_c/\lambda \sqrt{q^2}$ to $x=1-\xi$. We obtain
\begin{eqnarray}
H_U^1&=&-128\GG(-1)+128\GG(0)-32\GG(1)+64\GL(-1)-64\GL(0)+32\GL(1),
  \nonumber\\[7pt]
H_U^2&=&-32\xi\GG(-1)+32\xi\GG(0)+16\xi(2-\xi)\GL(-1)-32\xi\GL(0),
  \nonumber\\[7pt]
H_L^1&=&16(4+\xi)\GG(-1)-16(4+\xi)\GG(0)-8\xi(6-\xi)\GL(-1)
  +16\xi\GL(0)+8\xi\GL(1),\nonumber\\[7pt]
H_L^2&=&-16\xi\GG(-1)+16\xi\GG(0)+8\xi(2-\xi)\GL(-1)-16\xi\GL(0)
  -8\xi\GL(1),\\
H_F^4&=&-8(13-12\xi-\xi^2-8\Li(1-\xi))-48\ln\xi
  -64(1-\xi+\ln\xi)\ln\left(\frac{2\lambda}{v^2}\right),\nonumber\\[12pt]
H_U^{1x}&=&-128\GG(-1)+128\GG(0)+64\GL(-1)-64\GL(0),\nonumber\\[7pt]
H_U^{2x}&=&-32\xi\GG(-1)+32\xi\GG(0)+16\xi(2-\xi)\GL(-1)-32\xi\GL(0),
  \nonumber\\[7pt]
H_L^{1x}&=&16(4+\xi)\GG(-1)-16(4+\xi)\GG(0)-8\xi(6-\xi)\GL(-1)
  +16\xi\GL(0),\\[7pt]
H_L^{2x}&=&-16\xi\GG(-1)+16\xi\GG(0)+8\xi(2-\xi)\GL(-1)-16\xi\GL(0),
  \nonumber\\
H_F^{4x}&=&-64(2(1-\xi)-\Li(1-\xi))-64\ln\xi
  -64(1-\xi+\ln\xi)\ln\left(\frac{2\lambda}{v^2}\right),\nonumber\\[12pt]
H_U^{4z}&=&24(1-\xi)(3-\xi)+48\ln\xi,\nonumber\\[7pt]
H_L^{4z}&=&0,\\[7pt]
H_F^{1z}&=&-128\GG(0)+96\GG(1)+64\GL(0)-32\GL(1),\nonumber\\[7pt]
H_F^{2z}&=&0\nonumber
\end{eqnarray}
where
\begin{equation}
\GG(m):=\int_{2\lambda}^{1-\xi}x^{m-1}w_+(x)dx.
\end{equation}
Full analytical results for the integrals have been given before
in~\cite{gluon8}. Here we employ a small $\lambda$-expansion and retain
only the leading $\ln \lambda$ and the next-to-leading constant 
contributions in the small $\lambda$-expansion. One has 
\begin{eqnarray}
\GG(-1)&=&-\ln\left(\frac{1+v}{1-v}\right)
  -v\ln\left(\frac{\lambda\xi}{2v^2}\right),\\[12pt]
\GG(0)&=&v-\frac12\xi\ln\left(\frac{1+v}{1-v}\right),\\[12pt]
\GG(1)&=&\frac14(2+\xi)v
  -\frac\xi8(4-\xi)\ln\left(\frac{1+v}{1-v}\right),\\[12pt]
\GL(m)&:=&\int_{2\lambda}^{1-\xi}x^mt_+^\ell(x)dx,\\[12pt]
\GL(-1)&=&-\ln\left(\frac{1+v}{1-v}\right)
\ln\left(\frac{\lambda(1+v)^2}{2v^2}\right)
  -\Li\left(\frac{4v}{(1+v)^2}\right),\\[12pt]
\GL(0)&=&-v+\frac12(2-\xi)\ln\left(\frac{1+v}{1-v}\right),
  \\[12pt]
\GL(1)&=&-\frac38(2-\xi)v+\frac1{16}(8-8\xi+3\xi^2)
  \ln\left(\frac{1+v}{1-v}\right).
\end{eqnarray}
We have checked that the above leading and next-to-leading log expansion 
has sufficient accuracy for the numerical applications to be discussed in 
Sec.~4.

\section{Numerical results}
We are now in the position to discuss the various differential 
distributions of the polarization variables with regard to the gluon 
energy. The QCD dynamics is completely specified by the various hadron 
tensor components assembled in Eqs.~(\ref{eqn4})--(\ref{eqn5}). In order 
to exhibit the quark mass and flavour dependence of the gluon's 
polarization we shall present results for the top, bottom and charm pair 
production cases in some of the figures. We begin our discussion with the 
linear polarization of the gluon in the hadron event plane defined by 
the plane spanned by the quark, antiquark and the gluon. As discussed 
before the linear polarization of the gluon in the event plane is 
determined by the normalized $x$-component of the Stokes vector. To start 
with we integrate out the polar angle dependence in the numerator and 
denominator of the ratio of polarized and unpolarized cross sections given 
in Eq.~(\ref{eqn2}). One has
\begin{equation}
P^x(x)=\frac{d\sigma^x/dx}{d\sigma/dx},
\end{equation}
where $d\sigma^x/dx$ and $d\sigma/dx$ are the twice-integrated polarized
and unpolarized differential cross-sections appearing in Eq.~(\ref{eqn2}). 
In Fig.~2(a) we show the energy dependence of the linear polarization of 
the gluon produced in association with charm and top quark pairs at a 
c.m.\ energy of $\sqrt{q^2}=500~\GeV$. We have chosen to rescale the scaled 
gluon energy $x$ in this plot to its maximal value given by
$x_{\max}=1-\xi$. Fig.~2(a) shows that the linear polarization
remains close to its maximal value of 100\% at the soft gluon point over   
a large portion of gluon phase-space. For the top quark case the 
polarization remains above the value in the charm quark case when
plotted against $x/x_{\max}$\footnote{When the linear polarization is 
compared at equal gluon energies, this relation is reversed.}. In Fig.~2(b) 
we show the differential cross section which enters in the normalization of 
the polarization. The differential cross section for gluons produced in 
conjunction with top quark pairs can be seen to be approximately one order 
lower than for gluons produced in conjunction with charm quark pairs. 

In Fig.~3(a) we display the linear polarization of the gluon for three 
different values of the c.m.\ energy plotted against the scaled energy 
variable $x/x_{\max}$. As the lowest c.m.\ energy we choose 
$\sqrt{q^2}=370\GeV$ which is far enough above $t\bar t$-threshold for a 
perturbative treatment to be valid. The linear polarization remains quite 
close to its soft-gluon value of 100\% over most of the available 
phase-space due to the fact that there is not much energy available for 
the gluon so close to threshold. At the hard end of the spectrum the linear 
polarization has to go to zero for the simple reason that one can no longer 
define a hadronic plane in this collinear configuration. As the c.m.\ 
energy is increased, the linear polarization decreases for a given fixed 
fractional energy value $x/x_{\max}$. For the largest c.m.\ energy value 
$\sqrt{q^2}=1000\GeV$ in Fig.~3(a) the linear polarization is quite close to 
the zero mass limiting value also shown Fig.~3(a). As discussed before, 
the zero mass result is determined by the collinear configurations of the 
gluon, i.e. by the $\ln\xi$ contributions in Eq.~(\ref{eqn7}). Using the 
results of Eq.~(\ref{eqn7}), one obtains\footnote{The limiting value of 
the polarization agrees with the corresponding result Eqs.(4)--(9) 
in~\cite{gluon1} (second reference) in the limit $x_0,\delta\rightarrow 0$}. 
\begin{equation}
P^x(x)=\frac{2(1-x)}{2-2x+x^2}.
\end{equation}
Note that the zero mass linear polarization no longer depends on the 
electro-weak model parameters, on the flavour of the produced quark or
on the polarization of the beam. The electro-weak model dependence, the
flavour and beam polarization dependence is the same in the numerator and 
in the denominator and drops out when taking the ratio. It can be checked
that the electro-weak model, the flavour and beam polarization dependence
of the linear polarization is quite weak also in the massive case for the 
c.m.\ energies considered here. By comparing Fig.~2(a) with Fig.~3(a) one 
sees that the linear polarization in the charm quark case is already 
quite close to the zero mass case indicating that the approach to the 
asymptotic value is quite fast. This also explains the closeness of the 
$\sqrt{q^2}=1000\GeV$ curve to the zero mass curve.

In Fig.~3(b) we show the differential cross section for 
$e^+e^-\rightarrow t\bar tG$ for the same three c.m.\ energies as in 
Fig.~3(a). The differential cross section decreases as the c.m.\ energy 
is increased. As expected, the decrease is governed by the usual $q^2$-factor 
in the differential cross section.

In Fig.~4 we show the circular polarization of the gluon for the same
three c.m.\ energies as in Fig.~3 including again the zero mass case for
up-type quarks. It is clear that we are employing flavour tagging since 
the circular polarization would be zero otherwise. The circular 
polarization is positive in all four cases and peaks toward the hard end 
of the gluon spectrum\footnote{Note that the positive value holds for 
$E_q\ge E_{\bar q}$ and becomes negative for $E_q\le E_{\bar q}$.}. The 
circular polarization is zero at the soft end of the spectrum which is 
simple to understand since the linear polarization already saturates the 
total polarization bound of 100\% at this point. The circular polarization 
remains quite small over most of the range of gluon energies except for 
the zero mass case. The limiting zero mass behaviour is again governed by 
the collinear configuration determined by the $\ln\xi$ contributions in 
Eq.~(\ref{eqn7}). One obtains 
\begin{equation}\label{eqn8}
P^z(x)=-\frac{g_{14}}{g_{11}}\frac{x(2-x)}{2-2x+x^2}
\end{equation}
Note that the $\sqrt{q^2}=1000\GeV$ curve in Fig.~4 is still a large way
away from the zero mass limit indicating that the approach to asymptotia
is much slower for the circular polarization than what was observed
for the linear polarization.

The $x$-dependent factor in the mass zero formula Eq.~(\ref{eqn8}) for 
the circular polarization starts from zero at the soft-gluon point and 
reaches unity at the hard end of the spectrum. Apart from the $x$-dependent 
factor the size of the circular polarization is governed by the ratio of 
electroweak factors $-g_{14}/g_{11}$. Above the top quark threshold the 
ratio $-g_{14}/g_{11}$ is only mildly $q^2$-dependent but is strongly 
flavour dependent. For example, for $\sqrt{q^2}=370$, $500$ and $1000\GeV$ 
one finds $0.273$, $0.265$ and $0.257$ for the up-type quarks, and $0.678$, 
$0.670$ and $0.657$ for down-type quarks, respectively. From the numerical 
values of the ratio $-g_{14}/g_{11}$ it is clear that one generally obtains
much larger values for the circular polarization of down-type quarks than 
for up-type quarks. In Fig.~4(b) we provide a plot of the $x$-dependence of 
the circular polarization for $\sqrt{q^2}=1000\GeV$ in the zero mass case 
for up-type and down-type quarks. The above numerical results of the ratio
$g_{14}/g_{11}$ can be clearly identified at the hard end of the spectrum.
For the sake of comparison we have also plotted the corresponding curve
for bottom quarks with $m_b=4.1\GeV$. Mass effects tend to reduce the
circular polarization showing again that the approach to the asymptotic
formula Eq.~(\ref{eqn8}) is quite slow. 

In Fig.~5(a) we display the beam polarization dependence of the circular
polarization assuming that the electrons are polarized (the beam 
polarization dependence of the linear polarization is quite weak). The 
circular polarization of the gluons can be seen to be strongly dependent 
on beam polarization effects. One observes a strong polarization transfer
effect from the helicity polarization of the electrons to the circular 
polarization of the gluon which can enhance the circular polarization of 
the gluon up to a factor of ten. In Fig.~5(b) we exhibit the beam polarization 
dependence of the differential cross section. The differential cross 
section is enhanced by about 50\% for negative electron helicities and 
reduced by about the same amount for positive electron helicities.

In Fig.~6(a) and Fig.~6(b) we show the $\cos\theta$-dependence of the
linear and circular polarization of the gluon where we remind the reader
that $\theta$ is the polar angle of the gluon relative to the electron
beam. The dependence is shown for a c.m.\ energy of $\sqrt{q^2}=500\GeV$
and for three different gluon energies. The $\cos\theta$-dependence
of the linear polarization is quite weak for all three gluon energies
but becomes somewhat larger as one moves away from the soft gluon point
where the $\cos\theta$-distribution is flat. In contrast to this, the 
circular polarization shows a strong asymmetry effect and turns from 
positive to negative values as $\cos \theta$ moves from $-1$ to $+1$.
The asymmetry effect becomes larger as the gluon becomes harder. As 
Fig.~6(c) shows, the pronounced asymmetry of the circular polarization 
results mainly from the numerator of the relevant polarization expression. 
The differential cross section shown in Fig.~6(c) does not possess a very 
pronounced $\cos\theta$-dependence.  

In Fig.~7(a) we show a plot of the average linear polarization of the gluon
as a function of the c.m.\ energy for three different cutoff values 
$E_c=\lambda\sqrt{q^2}=3$, $5$ and $10\GeV$. Gluon energies of this 
magnitude are sufficient to make the corresponding gluon jets detectable. 
The average linear polarization of the gluon rises steeply from threshold 
and quickly attains very high values above 95\%. The linear polarization 
becomes larger for smaller values of $E_c=\lambda\sqrt{q^2}$ and tends to 
100\% as the cutoff parameter tends to zero. The approach to the asymptotic 
value $P^x=100\%$ is, however, rather slow. We also show the average linear 
polarization of the gluon limit when the quark mass goes to zero for a 
cutoff energy of $3\GeV$. The average linear polarization is smaller than 
in the massive case. In Fig.~7(b) we show the cutoff dependence of the 
total cross section as functions of the c.m.\ energy. The cross sections 
rise steeply from threshold and then turn over to their canonical 
$1/q^2$-behaviour. The cutoff dependence of the cross section is still 
moderate for the three chosen cutoff values. 

\section{Linear polarization in the gluon-beam plane}
In Secs.~2, 3 and~4 the linear polarization of the gluon has been 
specified relative to the event plane. A measurement of the polarization 
would require the reconstruction of the event plane which may not always 
be simple. If is much easier to determine the gluon-beam plane since the 
beam coordinates are known a priori. In order to determine the 
(unnormalized) components of the Stokes vector in the gluon-beam plane 
one needs to rotate the event plane Stokes vector around the gluon 
momentum axis by the azimuthal angle $\chi$ between the the two planes. 
One obtains\footnote{Note that the Stokes vector is not a true vector and 
thus its transformation behaviour under rotations differs from that of a 
true three-vector.}
\begin{eqnarray}\label{eqn9}
{H^x}'&=&\cos 2\chi H^x-\sin 2\chi H^y\nonumber\\
{H^y}'&=&\sin 2\chi H^x+\cos 2\chi H^y\\
{H^z}'&=&H^z\nonumber
\end{eqnarray}
where $({H^x}',{H^y}',{H^z}')$ are the components of the Stokes vector 
in the gluon-beam plane. Since one is rotating around the gluon axis, the 
circular component $H^z$ is not affected by this rotation. We have dropped 
all further indices on the hadron tensor in Eq.~(\ref{eqn9}).

We shall not write down the full polar $\cos\theta$- and azimuthal 
$\chi$-dependence of the polarized cross section in the new system but 
shall rather completely integrate out the polar and azimuthal angle 
dependence. One then obtains
\begin{eqnarray}
{H_T^x}'&=&-\frac12(H_T^x+H_4^y)\nonumber\\
{H_T^y}'&=&0
\end{eqnarray}
where the $y'$-component ${H^y}'$ drops out after azimuthal averaging. 
The hadron tensor components ``$T$'' and ``$4$'' on the right hand side are 
still specified in the event system and are given by (see~\cite{gluon9})
\begin{eqnarray}
H^x_T&=&\frac12(H_{+-}^x+H_{-+}^x)\ =\ \frac12(-H_{11}^x+H_{22}^x)\nonumber\\
H^y_4&=&-\frac12(H_{+-}^y-H_{-+}^y)\ =\ -\frac12(H_{12}^y+H_{21}^y).
\end{eqnarray}

Again these components can be calculated from the relevant tree-level
Feynman diagrams. One finds
\begin{eqnarray}\label{eqn10}
{H_T^{1x}}'(x,w)&=&\frac8{x^2}+8\xi t_+(x,w)
  -4\xi\frac{4-\xi}xt_+^\ell(x,w)\nonumber\\
{H_T^{2x}}'(x,w)&=&0.
\end{eqnarray}

Since we are interested in the energy dependence of the linear polarization, 
we also give their once-integrated forms. One has
\begin{eqnarray}
{H_T^{1x}}'(x)&=&8\left(\frac{2+\xi}x-\xi\right)\frac{w_+(x)}{x}
  -4\xi\frac{4-\xi}xt_+^\ell(x)\nonumber\\
{H_T^{2x}}'(x)&=&0.
\end{eqnarray}
It is quite evident that ${H_T^x}'$ has a smooth zero mass limit, i.e.\ 
there is no collinear singularity in the $(x,w)$-dependent structure 
function ${H_T^x}'(x,w)$ in Eq.~(\ref{eqn10}). This must be contrasted with 
the corresponding structure function describing the linear polarization in 
the event plane which possesses a collinear singularity (see 
Eq.~(\ref{eqn7})). We therefore anticipate that the linear polarization in 
the gluon-beam plane will be generally smaller than the linear polarization 
in the event plane and that it tends to zero in the zero mass limit.

We are now in the position to calculate the linear polarization of the 
gluon in the gluon-beam plane. The general expression for the linear
polarization in the gluon-beam plane reads
\begin{equation}
{P^x}'=\frac{g_{11}{H_T^{1x}}'+g_{12}{H_T^{2x}}'}
     {g_{11}(H_U^1+H_L^1)+g_{12}(H_U^1+H_L^2)}
\end{equation}
where we have retained the ${H_T^{2x}}'$ term in the numerator even though 
its contribution is zero.

In Fig.~8 we show the $(x/x_{max})$-dependence of the linear polarization
of the gluon in the gluon-beam plane for the top quark case for three
different c.m.\ energies. The linear polarization is generally small and
peaks toward the hard end of the spectrum. Also the linear polarization
tends to become larger when the c.m.\ energy is increased. This is 
counterintuitive from what was said before about the absence of a collinear 
singularity for the polarized structure function when $v\rightarrow 1$. 
The approach to the limiting behaviour of ${P^x}'$ (for $v\rightarrow 1$) 
is, however, so slow that one is still very far away from the limiting 
behaviour for the c.m.\ energies considered in Fig.~8. Note that the linear 
polarization neither vanishes at the soft end of the spectrum nor at the 
hard end. In fact, one obtains
\begin{equation}
{P^x}'(x=0)=\frac{g_{11}\left(-2(3-v^2)v+(1-v^2)(3+v^2)t_l\right)}
  {2((3+v^2)g_{11}+3(1-v^2)g_{12})\left(2v-(1+v^2)t_l\right)} 
\end{equation}
at the soft gluon point, where
\begin{equation}
v=\sqrt{1-\xi},\qquad t_l=\ln\left(\frac{1+v}{1-v}\right).
\end{equation}
In the zero mass limit one obtains ${P^x}'(0)\rightarrow 0$ as expected.
At the hard end of the spectrum for $x=x_{\rm max}=1-\xi$, the linear 
polarization is given by
\begin{equation}\label{eqn11}
{P^x}'(x_{\rm max})=\frac{g_{11}}{(3-v^2)g_{11}-(1-v^2)g_{12}}.
\end{equation}
Note that the mass zero limit of Eq.~(\ref{eqn11}) is finite, i.e.\ 
${P^x}'(x_{\rm max})\rightarrow 1/2$ as $v\rightarrow 1$. Again one would 
naively expect ${P^x}'\rightarrow 0$ in this limit. However, one is picking 
up a phase space zero in the denominator of the polarization expression at 
the hard gluon point which cancels the collinear singularity. The overall 
effect of the cancellation is that the polarization tends to a finite value 
as $v\rightarrow 0$. In fact, when one plots the energy dependence of the 
linear polarization for very small quark masses, the linear polarization 
tends to zero over the whole energy range, but shoots up to the value $0.5$ 
very close to the endpoint.

\section{Summary and conclusions}
We have provided a detailed discussion of the linear and circular gluon
polarization of gluons produced in association with heavy and light 
quark pairs in $e^+ e^-$ annihilations. We have studied beam polarization
and polar orientation effects on the polarization of the gluon. The linear
polarization of the gluon remains close to its classical soft gluon value
of 100\% and is only mildly dependent on the polarization of the beam, on 
the polar orientation of the gluon and on the flavour of the heavy quark 
pair produced in association with the gluon. Quark mass effects enhance 
the linear polarization of the gluon when compared at the same scaled enery 
scaled to the maximal energy. The circular polarization of the gluon is 
strongly dependent on the polarization of the beam, on the polar 
orientation of the gluon and on the flavour of the heavy quark pair produced 
in association with the gluon. The dependence on these effects can be 
exploited to optimally tune the circular polarization of the gluon. However, 
to see a nonzero circular polarization of the gluon one needs to tag the
flavour of the associated quark/antiquark since the circular polarization
is odd under charge conjugation and averages to zero without flavour 
tagging. This should not be so difficult for the heavy-flavoured quarks.
To actually measure the circular polarization of gluons one needs to
study the fragmentation of circularily polarized gluons into polarized
particles whose polarization can be measured. 

If one aims to study gluon polarization effects in the splitting process 
$e^+e^-\rightarrow t\bar tG(G\rightarrow GG,q\bar q)$, the present on-shell 
calculation should be sufficient to identify and discuss the leading 
effects of gluon polarization without having to perform a full $O(\alpha_s^2)$ 
calculation of $e^+e^-\rightarrow t\bar tGG$ and 
$e^+e^-\rightarrow t\bar t q\bar q$~\cite{gluon5}.

\vspace{1truecm}\noindent
{\bf Acknowledgements:} This work is partially supported by the BMBF, FRG, 
under contract No.~06MZ566, and by HUCAM, EU, under contract 
No.~CHRX-CT94-0579. The work of J.A.L. is supported by the DAAD, FRG.

\newpage

\vspace{1cm}
\centerline{\Large\bf Figure Captions}
\vspace{.5cm}
\newcounter{fig}
\begin{list}{\bf\rm Fig.\ \arabic{fig}:}{\usecounter{fig}
\labelwidth1.6cm\leftmargin2.5cm\labelsep.4cm\itemsep0ex plus.2ex}

\item Quark-gluon opening angle distribution for soft gluons illustrating 
the ``dead-cone'' effect\\
(a) opening angle distribution at a fixed c.m.\ energy of 
$\sqrt{q^2}=500\GeV$ for gluon energies of $0.5$, $1$, $2$, and $3\GeV$\\
(b) opening angle distribution for a fixed gluon energy of $6\GeV$ at three 
different c.m.\ energies of $370$, $500$, and $1000\GeV$

\item
(a) energy dependence of the linear polarization of the gluon in the event 
plane\\
(b) energy dependence of the differential cross section 
$e^+e^-\rightarrow t\bar tG$ and $e^+e^-\rightarrow c\bar cG$

\item
(a) energy dependence of the linear polarization of the gluon in the event 
plane for different c.m.\ energies\\
(b) energy dependence of the differential cross section for 
$e^+e^-\rightarrow t\bar tG$ for different c.m.\ energies

\item
(a) energy dependence of the circular polarization of the gluon produced 
with a top quark pair and a zero mass up-type quark pair for different 
c.m.\ energies\\
(b) energy dependence of the circular polarization of the gluon produced 
with zero mass up-type and down-type quarks and with a bottom quark pair at 
$\sqrt{q^2}=1000\GeV$

\item
(a) energy dependence of the circular polarization of the gluon in 
$e^+e^-(h_-)\rightarrow t\bar tG$ with electron beam polarization\\
(b) energy dependence of the differential cross section 
$e^+e^-(h_-)\rightarrow t\bar tG$ with electron beam polarization

\item
(a) polar angle dependence of the linear polarization of the gluon in the 
event plane\\
(b) polar angle dependence of the circluar polarization of the gluon\\
(c) polar angle dependence of the differential cross section

\item
(a) Mean linear polarization of the gluon in the event plane plotted 
against the c.m.\ energy for different cutoff energies\\
(b) total cross section for $e^+e^-\rightarrow t\bar tG$ plotted against 
the c.m.\ energy for different cutoff energies

\item Energy dependence of the linear polarization of the gluon in the 
gluon-beam plane
\end{list}
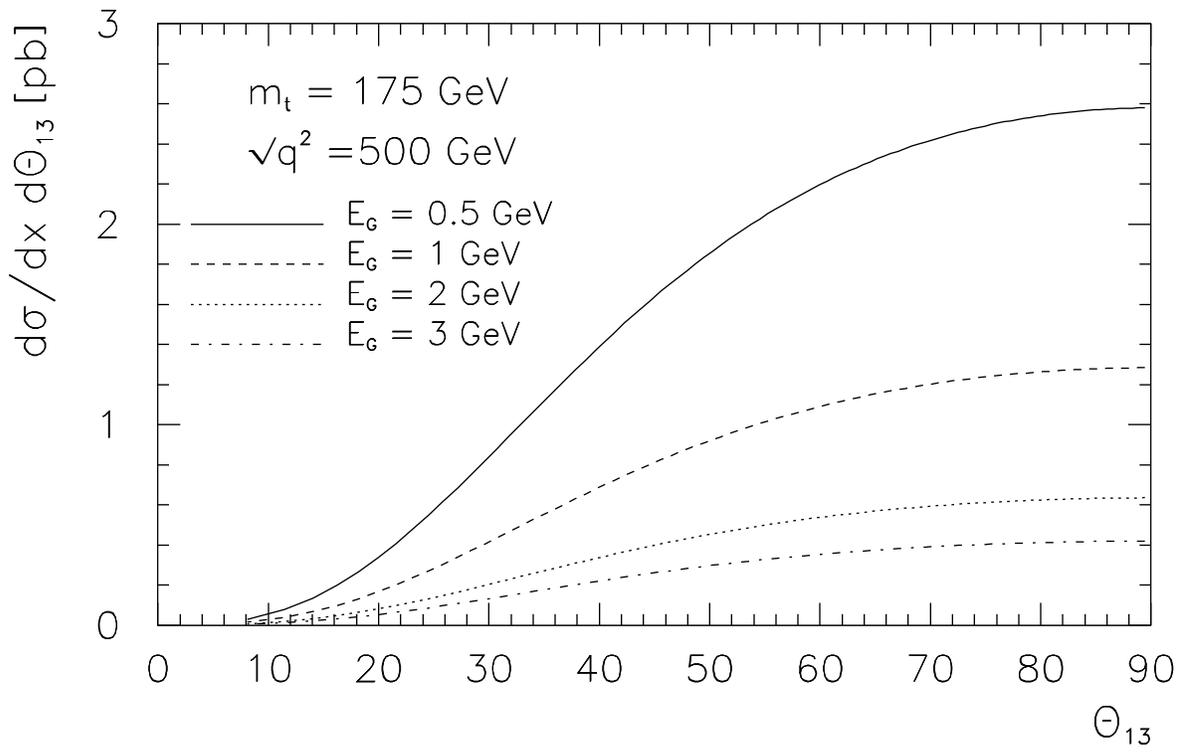
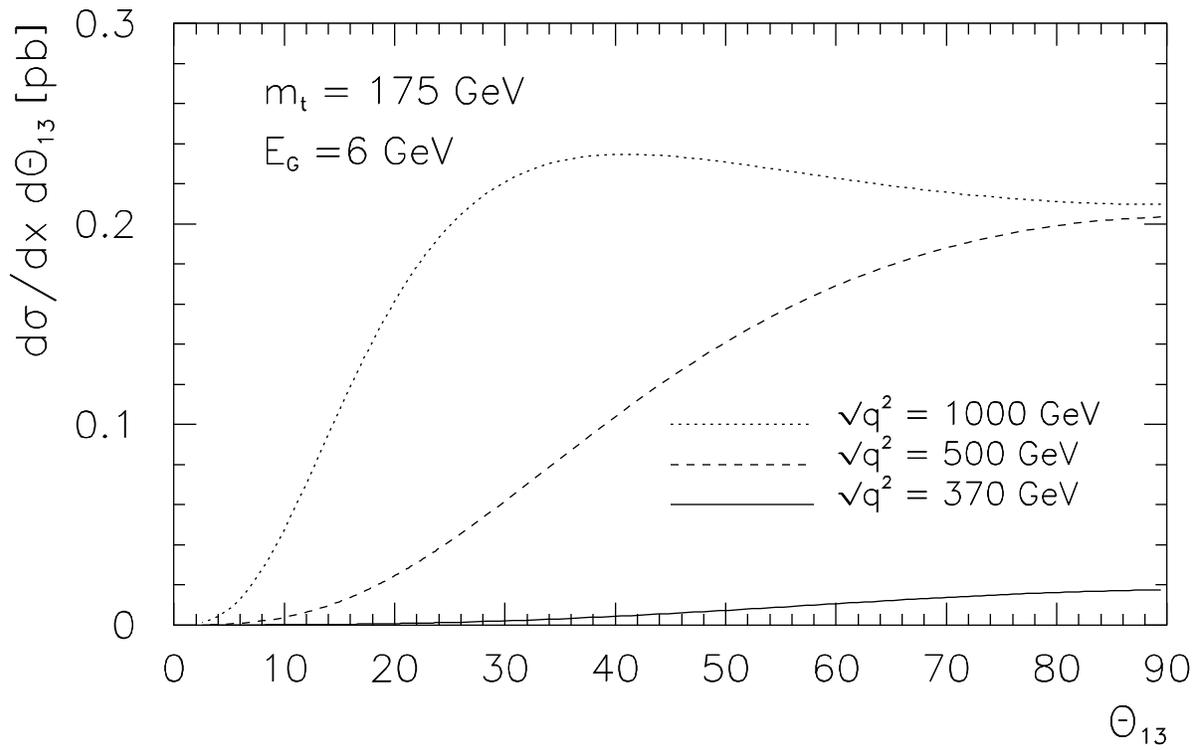
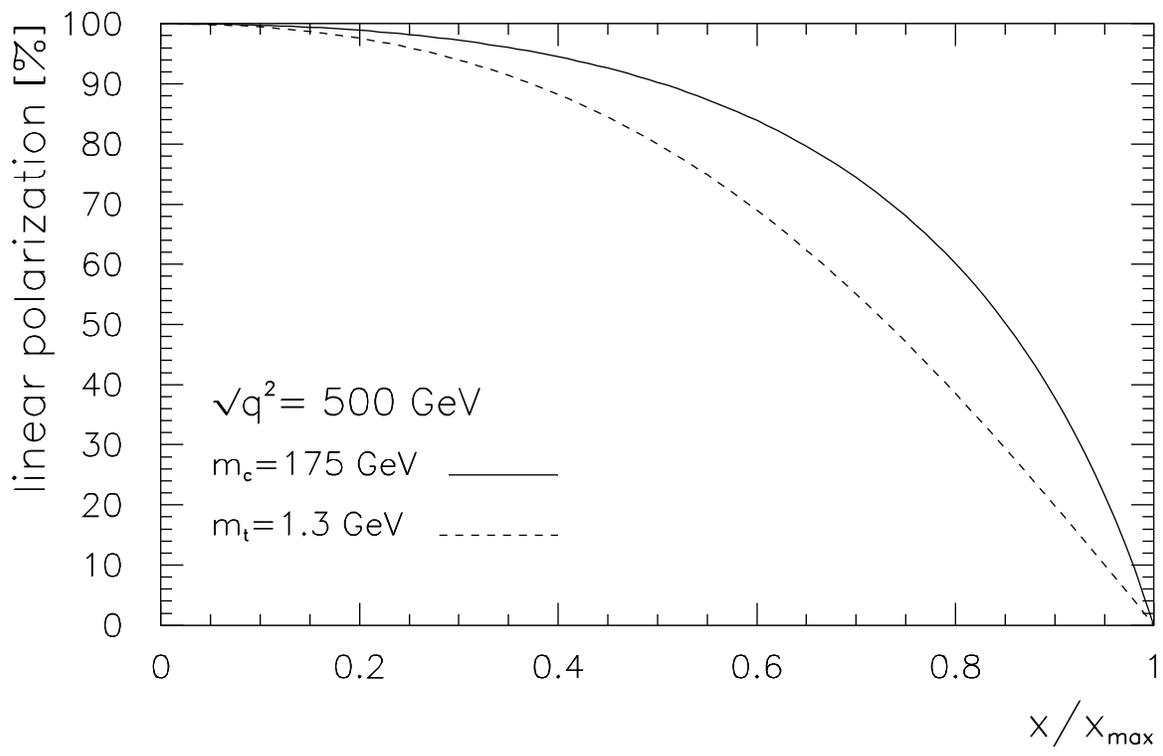
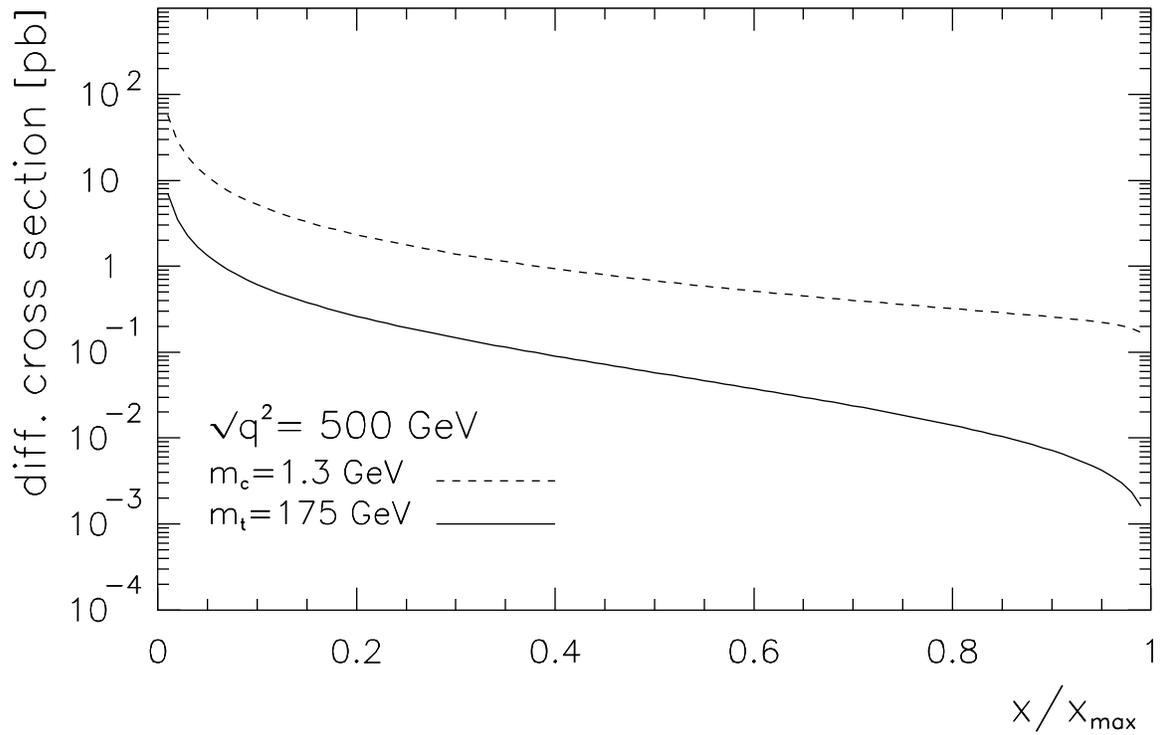
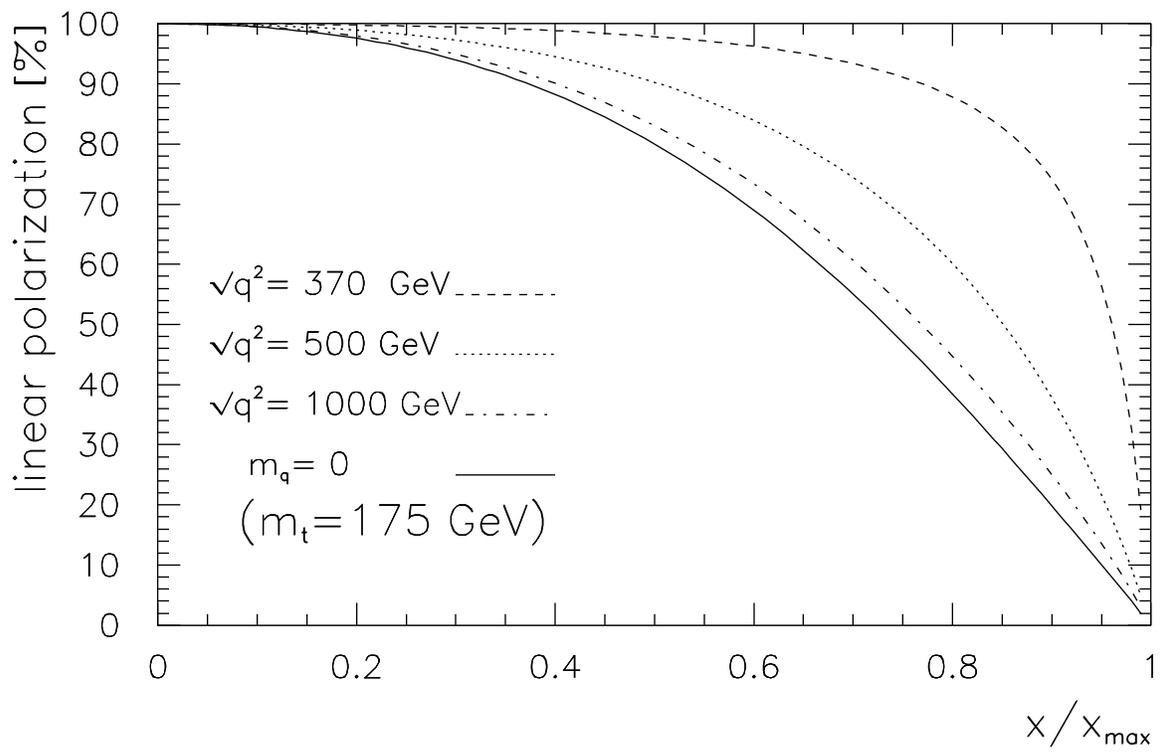
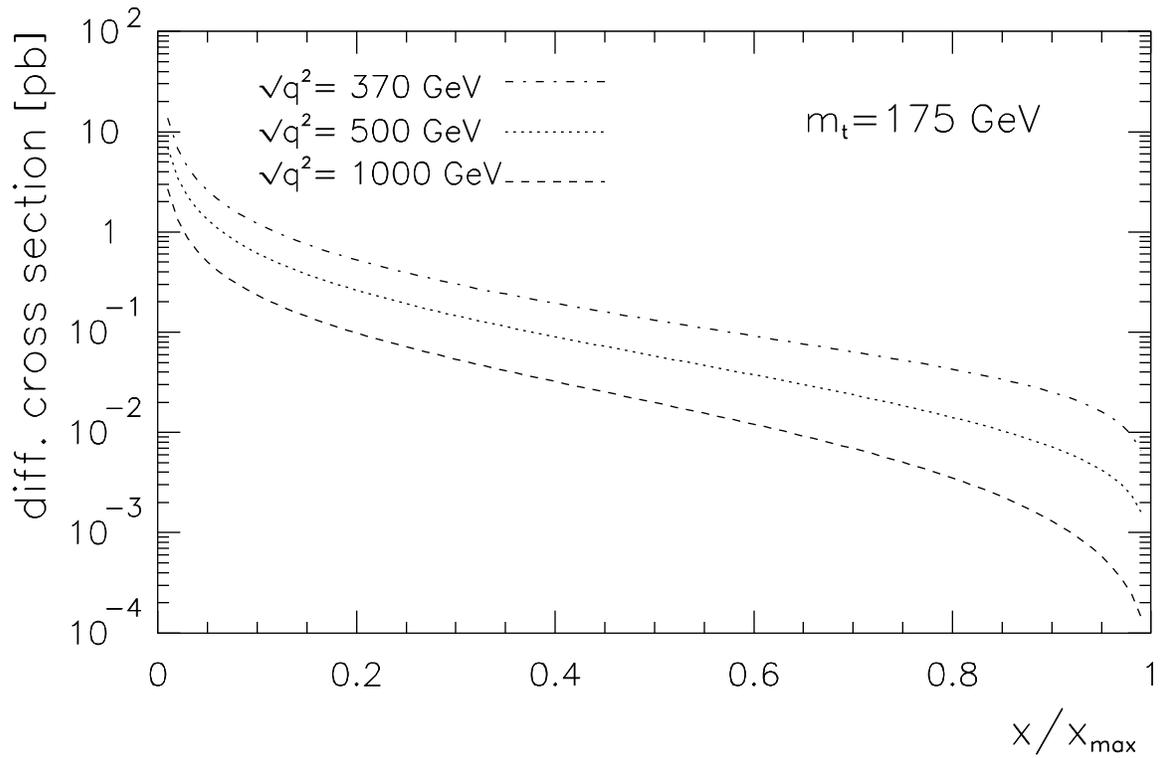
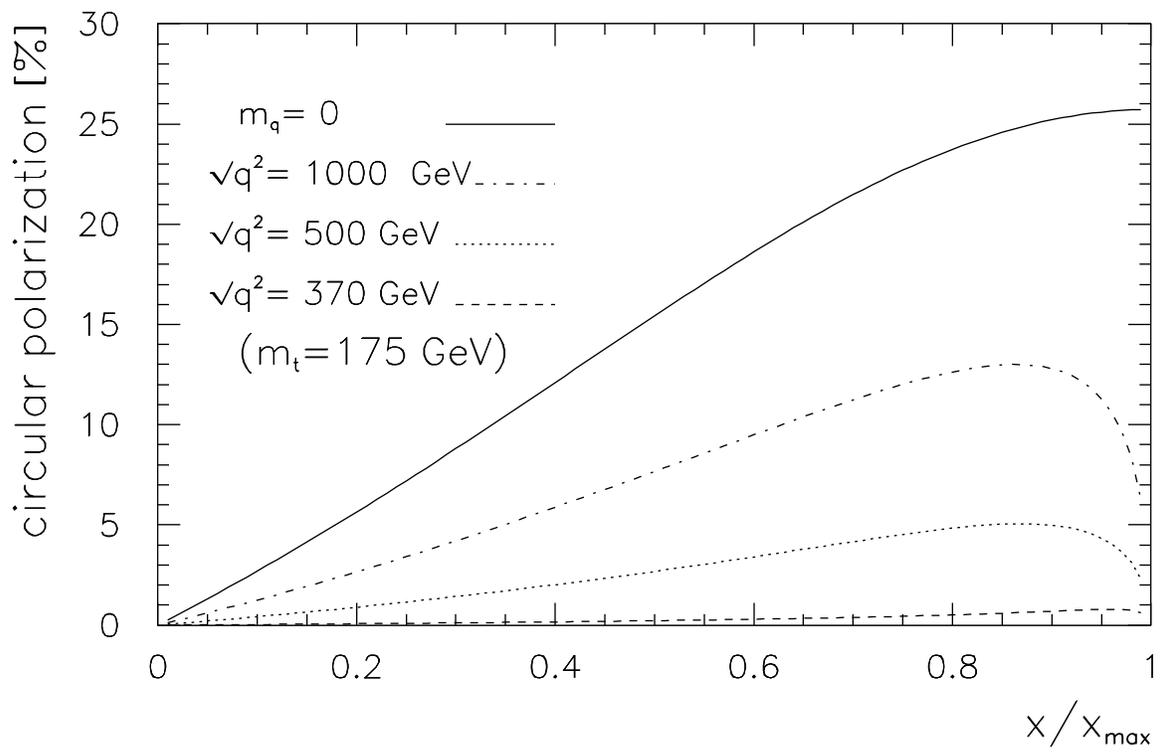
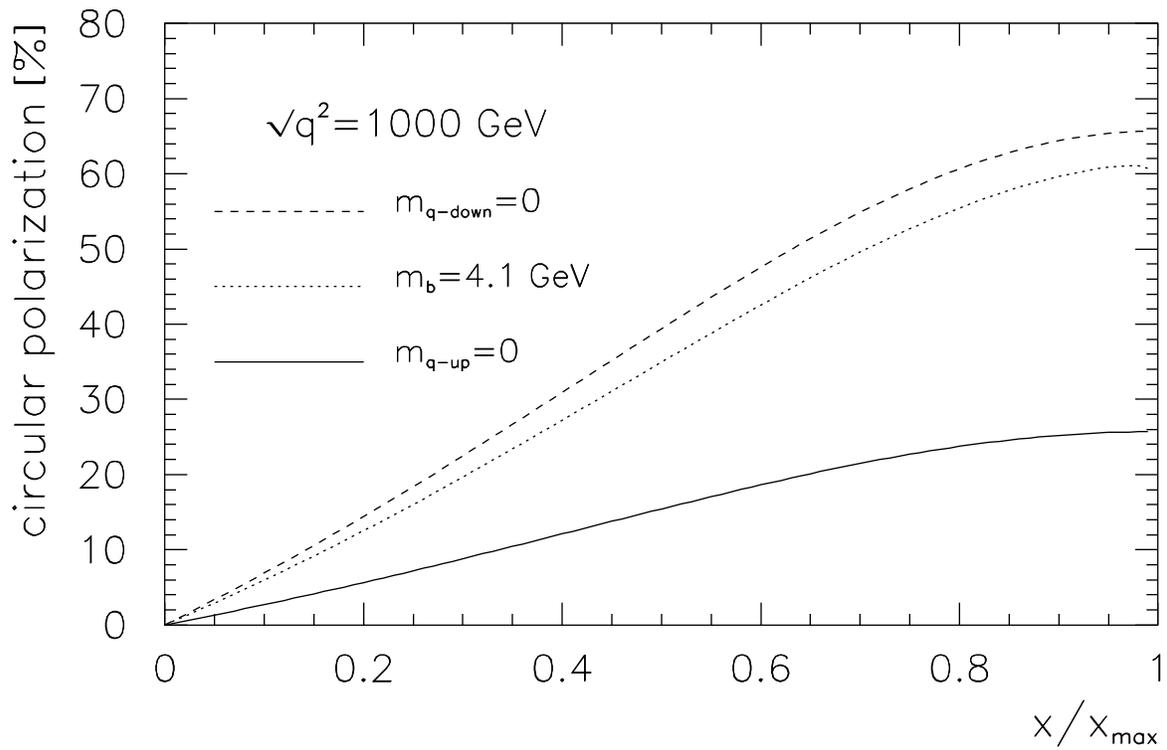
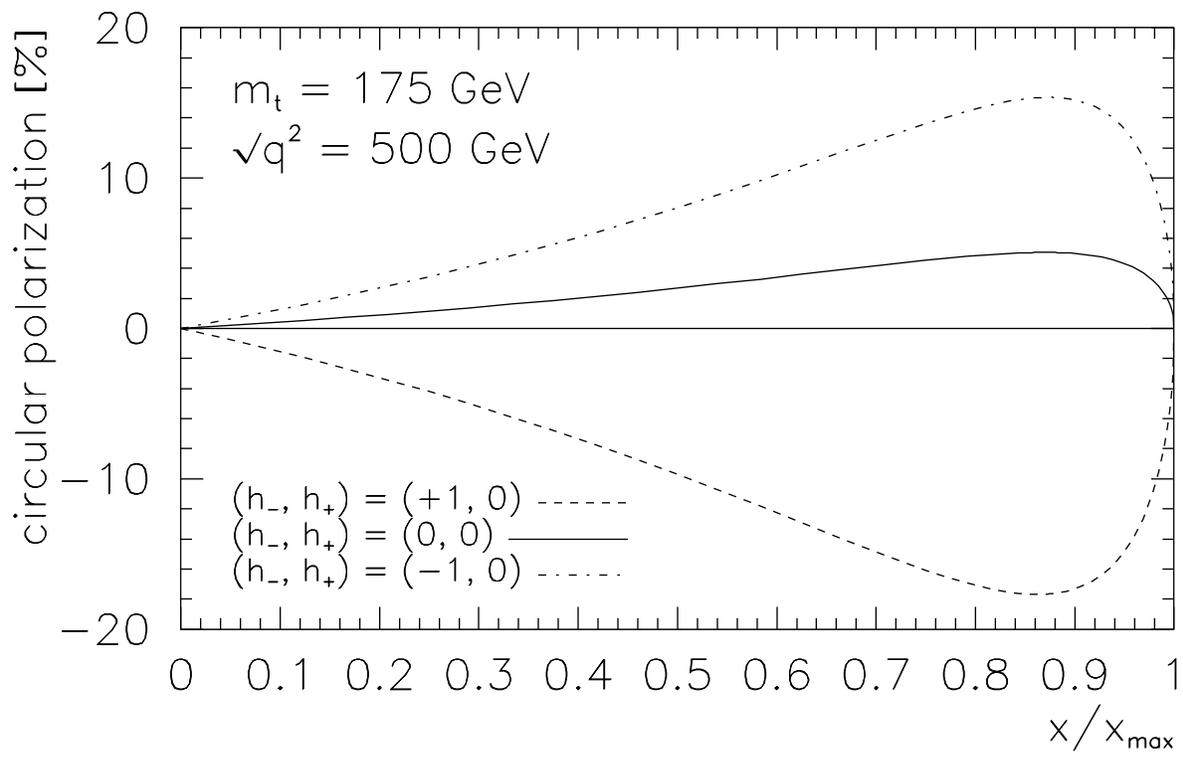
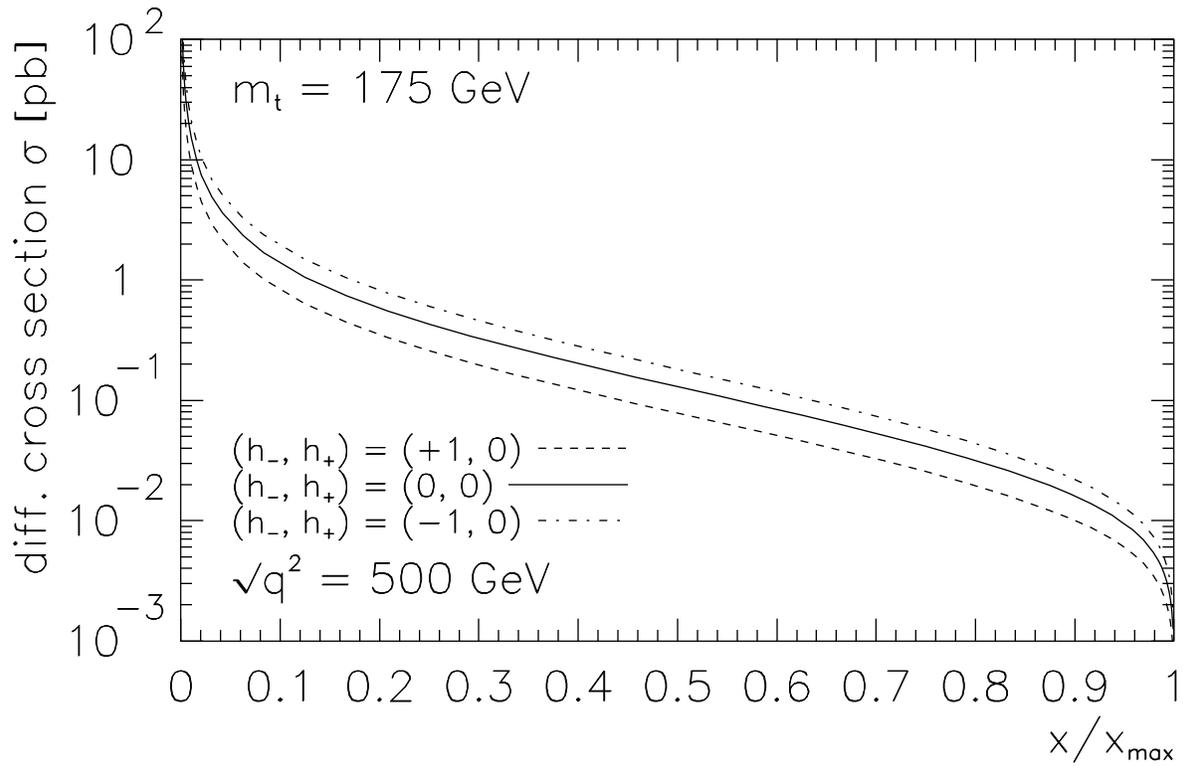
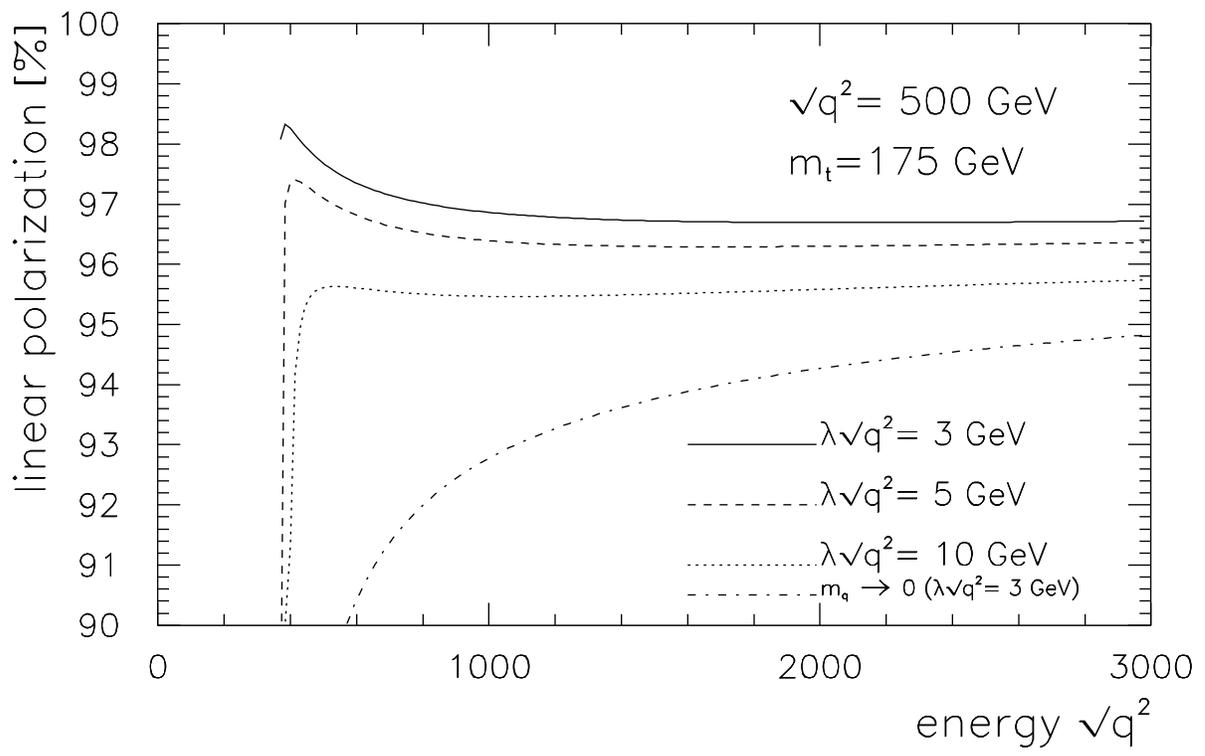
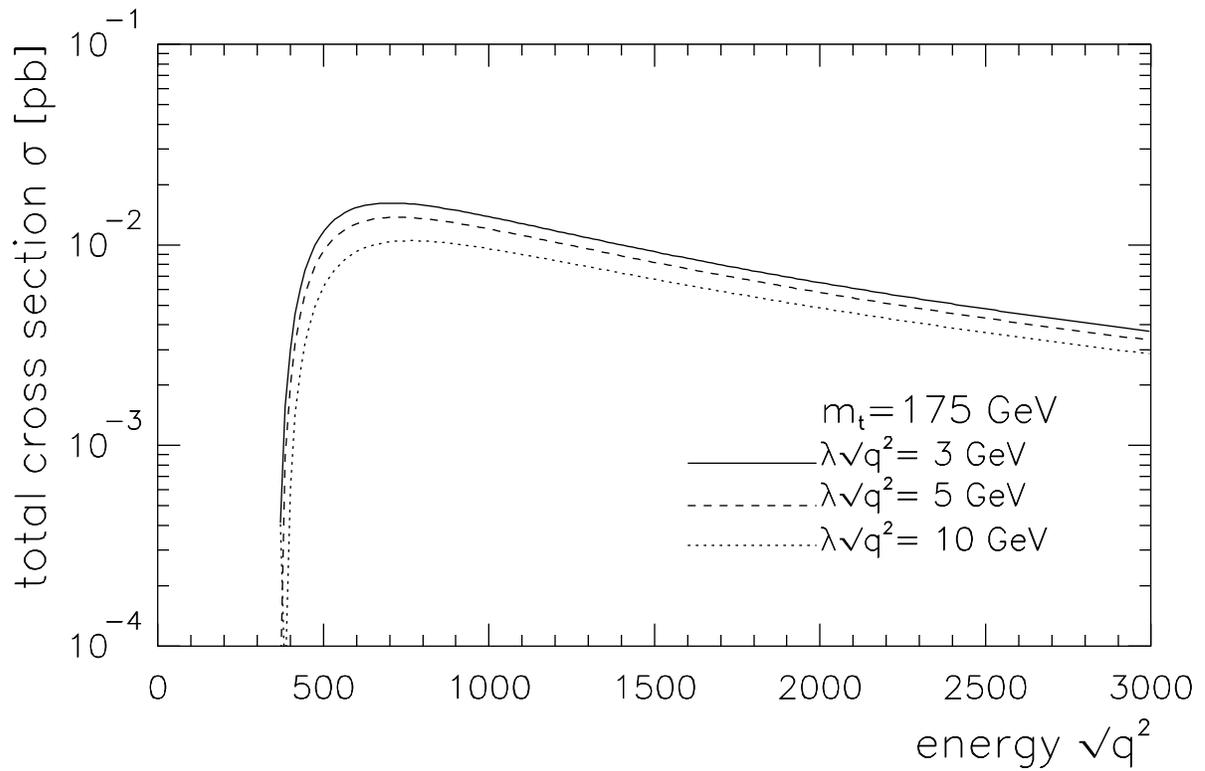
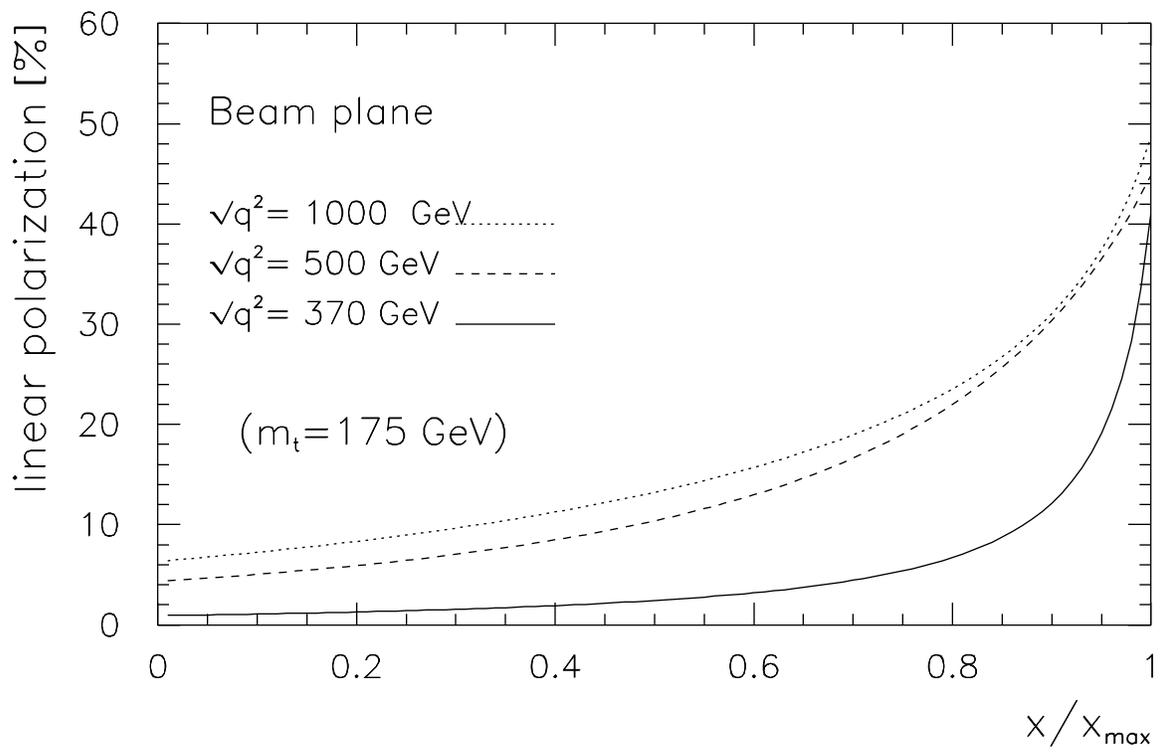
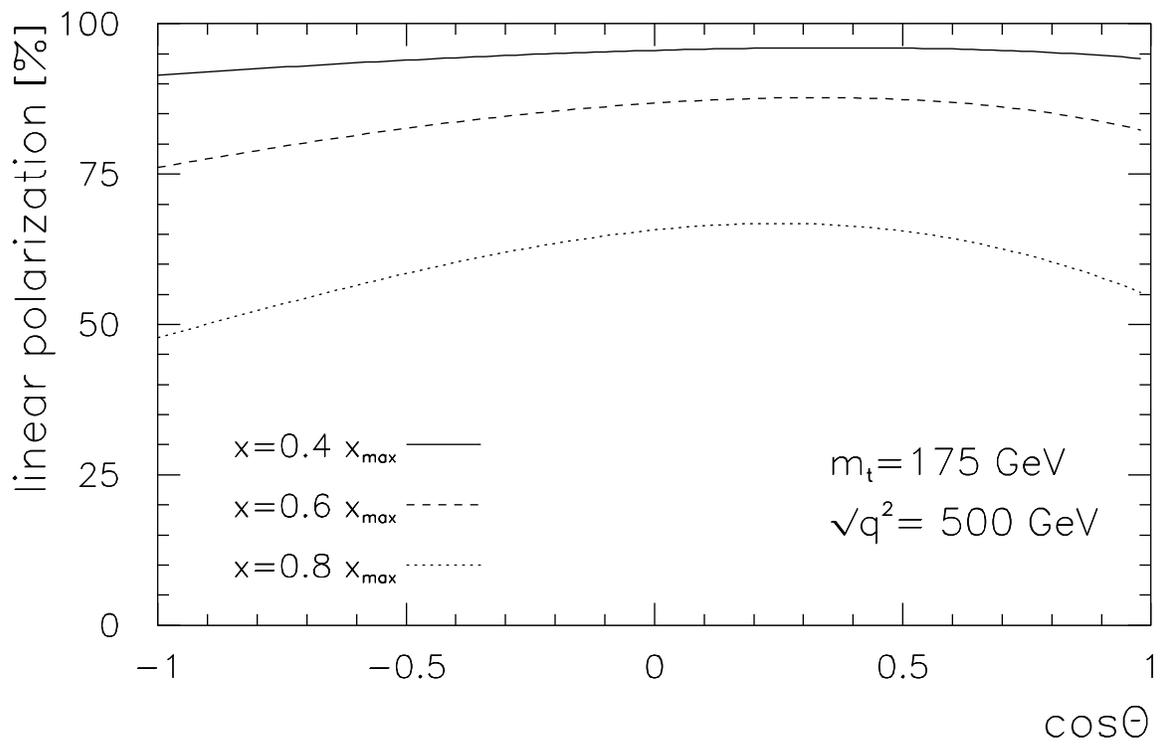
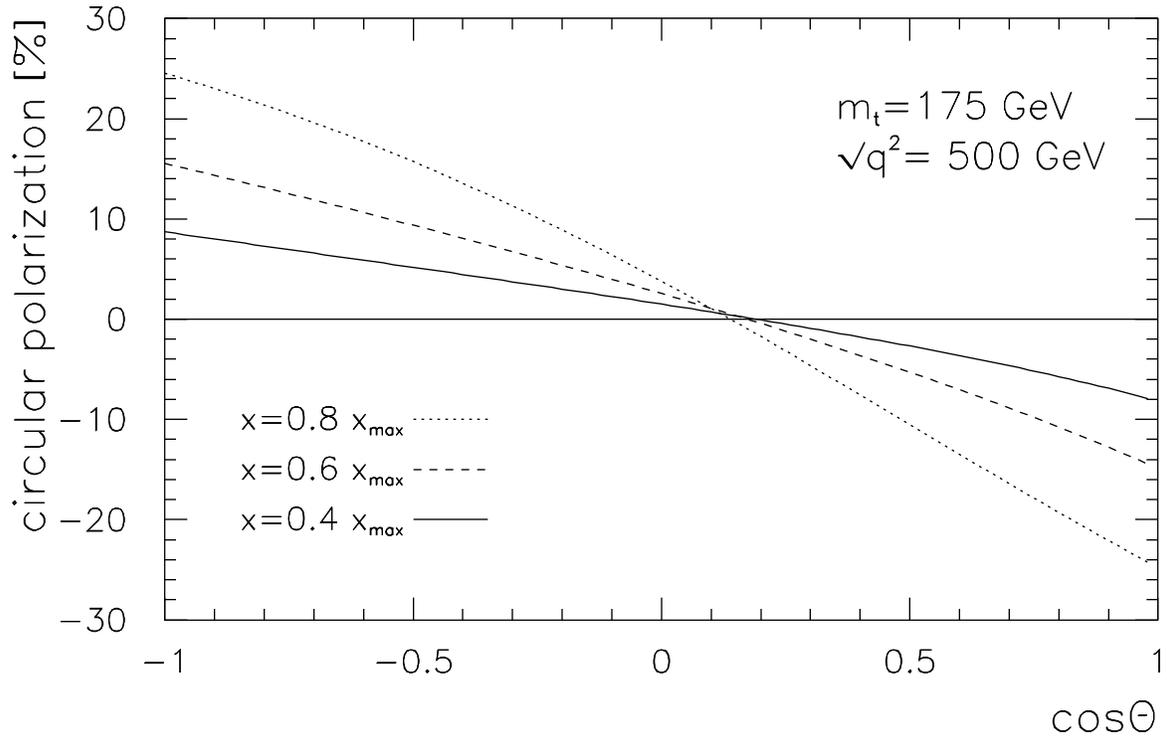
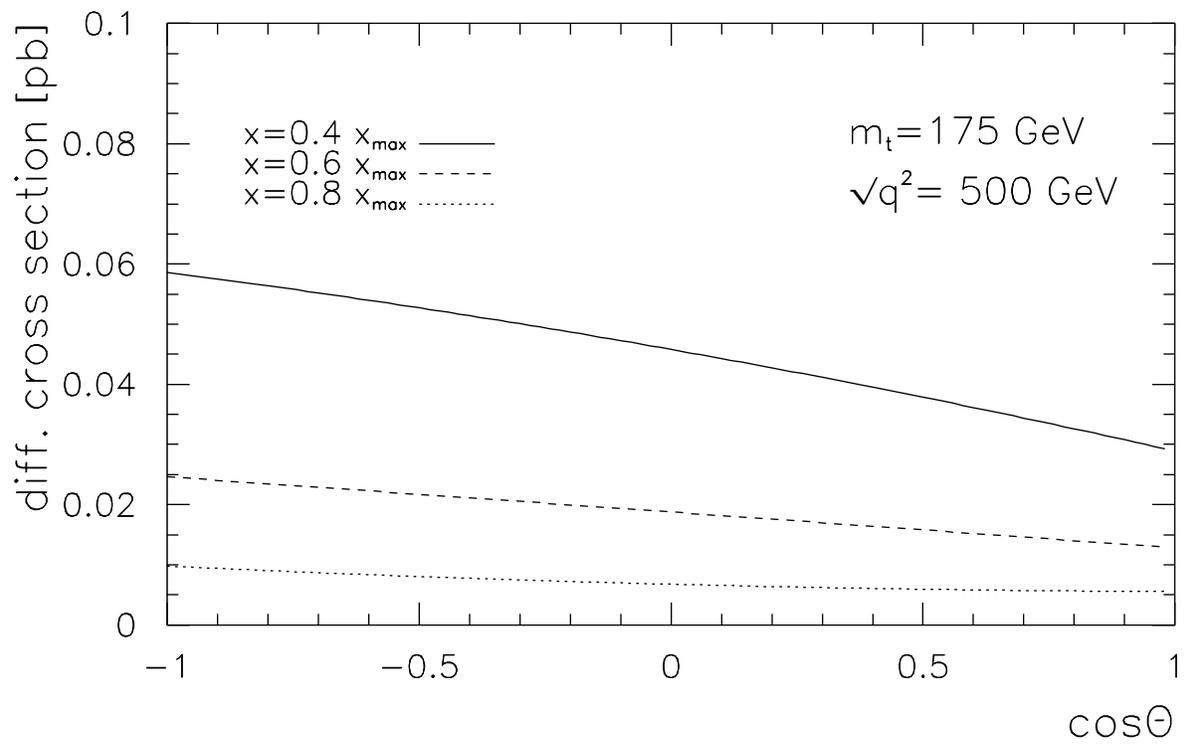
\end{document}